\documentclass[a4paper]{article}

\usepackage{INTERSPEECH2020,amsmath,graphicx,multirow,cite,url,booktabs}

\title{Interactive Text-to-Speech System via Joint Style Analysis}
  
\name{Yang Gao$^{1, \star}$ \thanks{$\star$ Work performed during internship at Facebook AI}, Weiyi Zheng$^{2}$, Zhaojun Yang$^{2}$, Thilo K{\"o}hler$^{2}$, Christian Fuegen$^{2}$, Qing He$^{2,}$$^\dagger$ \thanks{$\dagger$ Corresponding author}}

\address{$^{1}$Carnegie Mellon University \\
         \quad $^{2}$Facebook AI}
  
\email{yanggao@andrew.cmu.edu, \{wyz,zhaojuny,tkoehler,fuegen,qinghe\}@fb.com}

\begin{document}

\maketitle
\begin{abstract}

While modern TTS technologies have made significant advancements in audio quality, there is still a lack of behavior naturalness compared to conversing with people. We propose a style-embedded TTS system that generates styled responses based on the speech query style. To achieve this, the system includes a style extraction model that extracts a style embedding from the speech query, which is then used by the TTS to produce a matching response. We faced two main challenges: 1) only a small portion of the TTS training dataset has style labels, which is needed to train a multi-style TTS that respects different style embeddings during inference. 2) The TTS system and the style extraction model have disjoint training datasets. We need consistent style labels across these two datasets so that the TTS can learn to respect the labels produced by the style extraction model during inference. To solve these, we adopted a semi-supervised approach that uses the style extraction model to create style labels for the TTS dataset and applied transfer learning to learn the style embedding jointly. Our experiment results show user preference for the styled TTS responses and demonstrate the style-embedded TTS system's capability of mimicking the speech query style.

\end{abstract}

\section{Introduction}

With increasing interest in interactive speech systems such as voice assistants, there is an increased demand for human-like text-to-speech (TTS) systems. While recent technology advancements in speech synthesis have achieved human-like audio quality \cite{wavenet, alignT, attentionA, tacotron2}, the TTS's speaking style does not mimic the naturalness and expressiveness as in human conversations, because conventional speech interfaces respond to input speech queries with default speaking style learned from the TTS training dataset. To make the TTS more interactive, the TTS's response should vary  depending on the context and the speaking style of the input speech query. For example, when the user is speaking fast and rushing out the door in the morning, the TTS would match the hurried pace; and when the user is in a quiet place and is speaking softly, the TTS would respond with a soft and quiet voice. By detecting the input speech query's style and generating response accordingly, TTS can provide a more natural and customized user experience. One way to achieve this interaction is to incorporate two key components: a style extraction model that detects the speaking style of the input speech query and generates a style embedding, and a multi-style TTS system that can synthesize styled speech with respect to different style embedding inputs. 


The challenge lies in jointly training the style extraction model and the multi-style TTS so that the style embeddings generated by the style extraction model can be genuinely respected by the TTS, even though the two components are trained with different datasets and style labels. In this paper, the TTS dataset is a commissioned dataset recorded with professional voice talents. Only a small part of the TTS dataset has style labels. For the style extraction model, we make use of the external IEMOCAP dataset. These two datasets have different style labels. In order to achieve consistent labels between TTS training data and unseen queries, we incorporated both IEMOCAP dataset and a small portion of the labeled TTS dataset in the style classifier model's training. 

We first train a multi-modal style classifier using the IEMOCAP dataset with the model described in \cite{IE2018learning}. Taking the softmax layer of the style classifier as style embedding, the classifier serves as a style embedding extraction model. This model is applied to the unlabelled TTS dataset to generate the style embeddings in a semi-supervised fashion. By using the style embedding as additional auxiliary features for the TTS system, we could train a controllable multi-style TTS system that learns to respect given target styles. During speech synthesis, style embedding is first extracted from the input speech query and then fed into the TTS system to produce response in matching styles. In summary, we developed an interactive multi-style TTS system that could lead to natural, expressive human-machine speech interactions. The multi-style TTS system is evaluated using comprehensive subjective experiments. 

\section{Related work}
\subsection{Emotion recognition}
Early approaches on emotion recognition have mostly been inspired by psychology studies \cite{lee2011emotion, mower2010framework}. Recently, deep neural networks (DNNs) have first been used to learn high-level representations for utterance-level emotion recognition \cite{han2014speech}.  Trigeorgis \emph{et al.} further applied convolutional neural networks (CNNs) to model context-aware emotion-relevant features, which are then combined with long short-term memory (LSTM) networks aiming towards end-to-end emotion modeling \cite{trigeorgis2016adieu}. Fundamentally, the expression of emotions is usually conveyed through multi-modal behavior channels, including speech, language, body gestures, or facial expressions. Thus, emotion recognition is often formulated as a classification problem of utterances using these multi-modal signals. Reference \cite{Emotionmultimodal} proposed a multi-modal dual recurrent encoder to simultaneously model the dynamics of both text and audio signals within an utterance to predict emotion classes. This architecture has achieved state-of-the-art performance on IEMOCAP\cite{busso2008iemocap} dataset, which is a multi-modal emotion dataset and has been widely used in the affective computing community.

\subsection{Expressive TTS}
One popular topic in the recent research of TTS is expressive TTS. Expressive TTS has been studied for years from the HMM-based synthesis using style modeling with control vector \cite{HMMmodeling,tachibana2004hmm,yamagishi2004speaking} to the state-of-the-art prosody transfer expressive TTS work \cite{latentstylefactor, prosodytransfer, styletoken}, which is aiming at achieving controllable style synthesis in TTS. However, to learn and synthesize specific styles, there are limitations with unsupervised style factorization learning \cite{styletoken}. Since the disentanglement of different styles is heavily influenced by randomness and the choice of hyper-parameters \cite{ICMLbest2018}, the learning of specific target styles is not completely controllable. 

Under supervision with explicit prosody labels, the styles could be learned with direct guidance \cite{controlTTSwithlabeldata, Controllableprinciples, rabiee2019adjustingeTTS}. Supervised learning requires a large amount of labeled data, giving difficulties in the development of expressive TTS research and applications. Furthermore, the data labels for styles may not be well overlapped with the needs. An approach to tackle this is proposed in \cite{IE2018learning}. But, the external dataset and the synthesis dataset Blizzard 2017 \cite{king2017blizzard} have differences in background noise, recording environment, speech quality, etc. With the differences between these two datasets, the classifier trained using an external dataset may not be well-adapted to extract representations from the synthesis data. The final emotion synthesis accuracy is 41\% on four emotions \cite{IE2018learning} evaluated by listeners, which may be caused by the domain gap between the TTS dataset and the external dataset. 

\section{Datasets}

\subsection{TTS dataset}
\label{sssec:subsubhead}
The TTS dataset was recorded in voice production studio by multiple professional voice talents with $24$kHz sampling rate. It has balanced phonemic and textual information. After labelling to accommodate with the task, $7\%$ of the TTS dataset has utterance level style labels including happy, sad, neutral, angry, rushed, and soft. Details of the data are summarized in Table \ref{tab:data}. These utterances are used, as additional data, to train a multi-speaker style classifier, described in Section \ref{styleModel}. To train the multi-style TTS, we use 40,244 utterances from a single speaker which contains around 3000 style labelled utterances. The style embeddings for unlabelled portion are extracted using the style extraction model, more details in Section \ref{styleModel}.

\subsection{IEMOCAP dataset}
\label{sssec:subsubhead}
To compensate for the limited amount of labeled data in our TTS dataset, we chose IEMOCAP \cite{busso2008iemocap}, which is widely used for emotion recognition, to complement our training data. In this dataset, both video and audio were recorded from ten actors in dyadic sessions under scripted and spontaneous communication scenarios. The dataset contains 12.5 hours of recordings with a sampling rate of 22kHz. Each utterance contains one emotion label, such as neutral, happy, sad, anger, surprise, etc. To be consistent with former research \cite{IE2018learning, Emotionmultimodal} and also be suitable for our own interaction goal, we select the following emotions in our study: neutral, happy, sad, and angry. Similar to the approach in \cite{Emotionmultimodal}, we merge utterances with excited emotion with those of happy emotion.

\begin{table}[t]
\caption{Training data style label statistics}
\label{tab:data}
\vspace{-0.2cm}
\resizebox{\columnwidth}{!}
{
\begin{tabular}{lc | cccccc}
\toprule
Dataset & Split & \multicolumn{1}{c}{Rushed} & \multicolumn{1}{c}{Soft} & \multicolumn{1}{c}{Neutral} & \multicolumn{1}{c}{Happy} & \multicolumn{1}{c}{Angry} & \multicolumn{1}{c}{Sad}  \\
\hline
\midrule
TTS Dataset & \begin{tabular}{@{}c@{}@{}c@{}} Train \\ Dev \\ Test \\ All \end{tabular} 
& \begin{tabular}{@{}c@{}} 1145 \\105\\124 \\1374 \end{tabular} 
& \begin{tabular}{@{}c@{}} 1814 \\161 \\220 \\2195\end{tabular} 
& \begin{tabular}{@{}c@{}} 4481 \\ 439 \\506 \\5426 \end{tabular} 
& \begin{tabular}{@{}c@{}} 885 \\79 \\93 \\1057  \end{tabular} 
& \begin{tabular}{@{}c@{}} 140 \\ 13 \\17 \\170  \end{tabular} 
& \begin{tabular}{@{}c@{}} 35 \\ 3 \\2 \\40 \end{tabular} 

\\ 
\hline
\midrule
IEMOCAP & \begin{tabular}{@{}c@{}@{}c@{}} Train \\ Dev \\ Test \\ All \end{tabular} 
& \begin{tabular}{@{}c@{}} -- \\-- \\--\\-- \end{tabular} 
& \begin{tabular}{@{}c@{}} -- \\--\\--\\-- \end{tabular} 
& \begin{tabular}{@{}c@{}} 1390\\ 100 \\218 \\1708 \end{tabular} 
& \begin{tabular}{@{}c@{}} 1307 \\90 \\239 \\1636  \end{tabular} 
& \begin{tabular}{@{}c@{}} 865 \\ 61 \\177 \\1103  \end{tabular} 
& \begin{tabular}{@{}c@{}} 883 \\ 62 \\139 \\1084 \end{tabular} \\

\bottomrule 

\end{tabular}

}

\end{table}

\section{Framework and Models}
\label{sec:pagestyle}
\label{sec:format}

\subsection{Semi-supervised style transfer learning}\label{styleModel}

\begin{figure}[t]
 \centering
 \includegraphics[width=\linewidth]{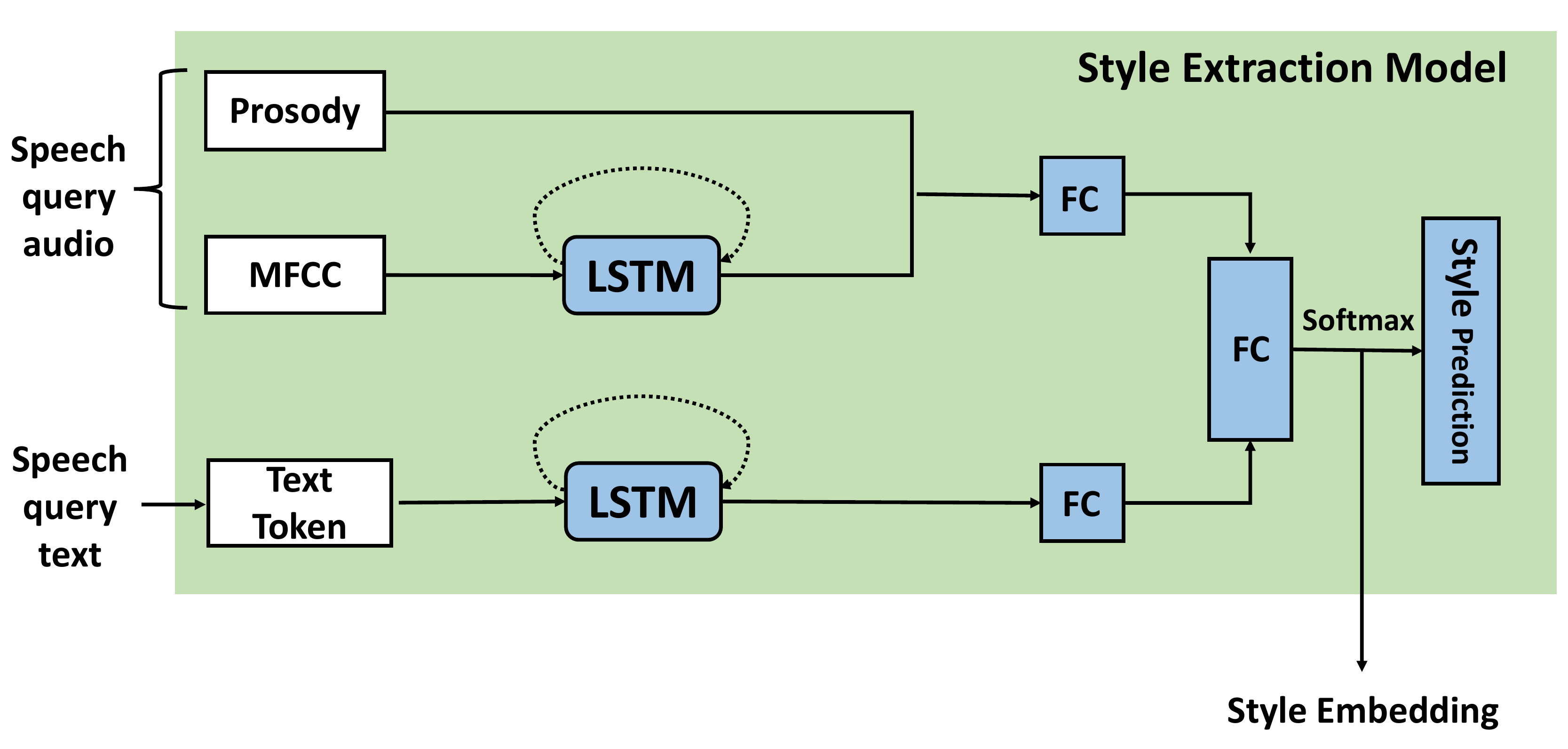}
 \vspace{-0.8cm}
 \caption{Multimodal style extraction model}
 \label{fig:Styleframework}
 \vspace{-0.3cm}
\end{figure}

For speech style classification, we used the multimodal dual recurrent encoder (MDRE) model adapted from \cite{Emotionmultimodal}. As shown in Figure \ref{fig:Styleframework}, the model is composed of two separate recurrent encoders for audio and text modeling, respectively. The audio model uses 39 dimensions Mel-frequency Cepstral Coefficients (MFCC) features and utterance level prosody feature extracted using openSMILE \cite{openSMILE} as inputs, and the text model uses 300-dimension embeddings to represent each word token. The MFCC, prosody, and text features are the same as described in \cite{Emotionmultimodal}. The audio encoder output is concatenated with the text encoder output, then fed into a fully-connected layer to produce the final classification. We changed the loss function from sigmoid cross-entropy to softmax cross-entropy as it produced significantly better results for our training task. We use the softmax layer output as embedding features, which can be interpreted as a weighted representation of different speaking styles. The softmax feature as embedding is shown in Figure \ref{fig:Styleframework}.

The style classifier is used to generate style embedding from the speech query during inference, as well as to extract style embedding for the TTS training dataset. At first, we trained the style classifier using the IEMOCAP dataset and applied it to generate style features on the TTS dataset. However, the classifier gives inaccurate predictions on the TTS dataset due to domain mismatch between the TTS dataset and the IEMOCAP dataset. Therefore, we labeled a small part of our TTS dataset using one label per utterance as Table \ref{tab:data} and fine-tuned the style classifier using these labels as in Section \ref{ssec:stylec}.

\subsection{Multi-style TTS system}

\begin{figure}[t]
 \centering
 \includegraphics[width=\linewidth]{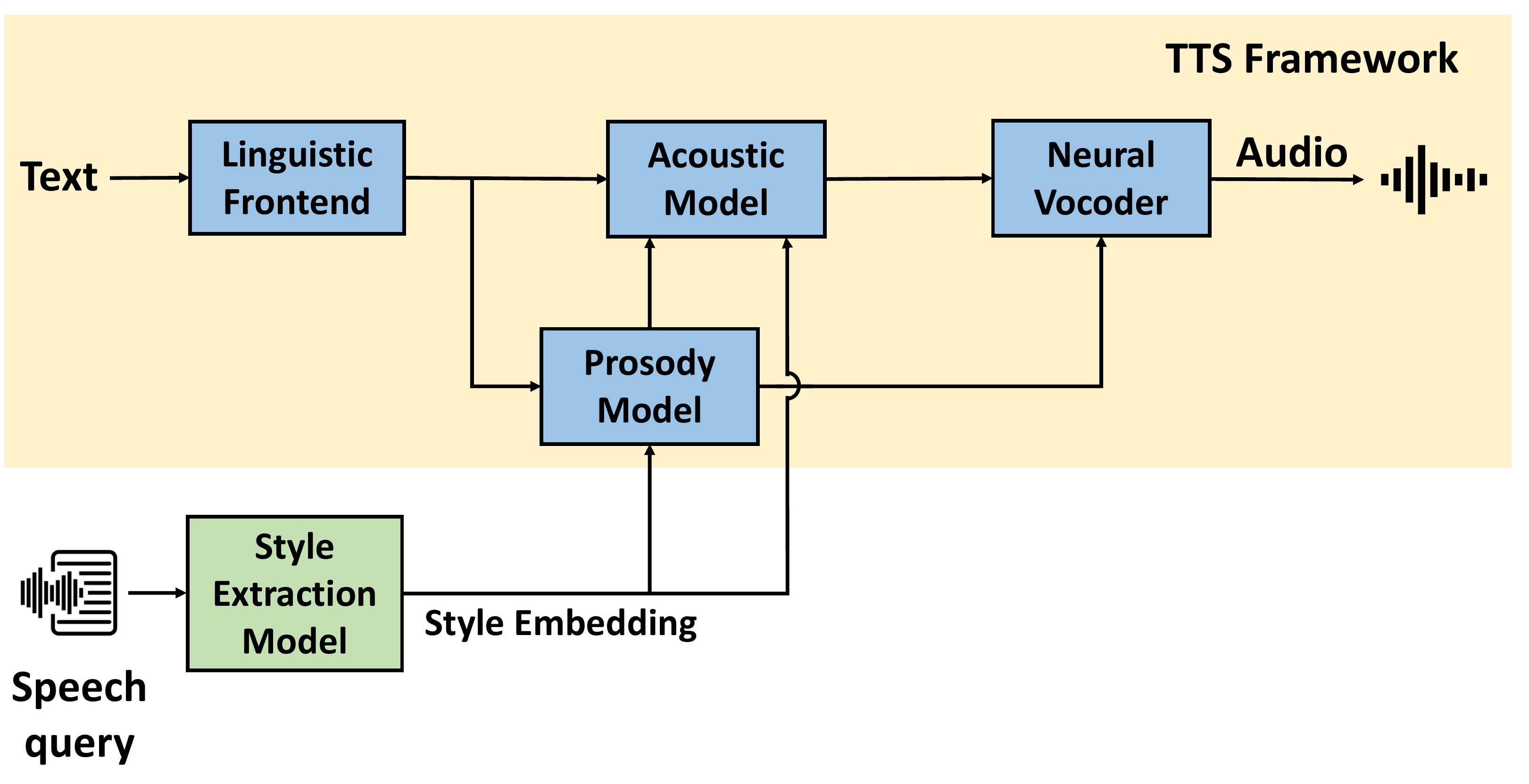}
 \caption{Style embedded TTS framework: the style extraction model generates the style embedding based on the user speech query (text + audio), which is used to condition the TTS synthesis.}
 \label{fig:framework}
 \vspace{-0.7cm}
\end{figure}

Figure \ref{fig:framework} shows the architecture of the expressive TTS system. It consists of a style embedding extraction component that generates the style embedding from speech query and a multi-style TTS, which uses the style embedding to synthesize its response in matching style. 
As shown in Figure \ref{fig:framework}, our TTS pipeline is a multi-model framework that consists of a linguistic frontend, a prosody model, an acoustic model, and a conditional neural vocoder. Specifically, the input text is first converted to linguistic features through a text normalization component followed by a joint-sequence grapheme to phoneme model. Then, the linguistic features, along with any conditional features such as style embedding, speaker IDs are used to produce the prosodic features such as duration and F$_{0}$. The prosody model consists of a single layer LSTM model with 256 hidden units with content-based global attention \cite{luong2015effective}, whose context vector contains linguistic features of the entire utterance. It is important to build a separate prosody model in the pipeline because it allows easier control for the speech style during synthesis time. Then, linguistic features combined with prosodic features are used to generate the 13-dim MFCC spectral acoustic features. The acoustic models consist of a two layer uni-directional LSTM with 256 hidden units per layer.  At the last stage, a conditional neural vocoder using the WaveRNN \cite{kalchbrenner2018efficient}, takes in the 13-dim MFCC along with the F0 feature to synthesize a 24kHz audio waveform. Our WaveRNN model consists of a single layer gated recurrent unit (GRU) with 1024 hidden units. The speaking style of the synthesized speech is controlled by the conditional style embedding feature, which can be pre-defined or extracted using the style extraction model from the input query, as in Figure \ref{fig:framework}.

\section{Experiments and results}
\label{sec:typestyle}

\begin{table*}[t]
\caption{Style classification on TTS data: AdaBN helps the domain adaption between IEMOCAP dataset and TTS dataset, improving the weighted accuracy of six style classes.}
\label{tab:results}
\vspace{-0.6cm}
\begin{center}
\scalebox{1}{

\begin{tabular}{lccccccccc}
\toprule

\multirow {2}{*}{Dataset} & \multirow{2}{*}{Trick} & \multirow{2}{*}{Neutral} & \multirow{2}{*}{Rushed} & \multirow{2}{*}{Soft} & \multirow{2}{*}{Happy} & \multirow{2}{*}{Angry} & \multirow{2}{*}{Sad} & \multicolumn{2}{c}{{\bf Accuracy}} \\
& & & & & & & & {Weighted} & {Unweighted} \\

\hline
\midrule
Train & \begin{tabular}{@{}c@{}} BN \\ AdaBN \end{tabular} 
& \begin{tabular}{@{}c@{}} 0.984 \\ 0.953 \end{tabular} 
& \begin{tabular}{@{}c@{}} 0.871 \\ 0.847   \end{tabular} 
& \begin{tabular}{@{}c@{}} 0.964 \\ 0.918 \end{tabular} 
& \begin{tabular}{@{}c@{}} 0.892 \\ 0.903 \end{tabular} 
& \begin{tabular}{@{}c@{}} 0.176 \\ 0.353 \end{tabular} 
& \begin{tabular}{@{}c@{}} 0.0 \\ 0.0 \end{tabular} 
& \begin{tabular}{@{}c@{}} 0.779 \\ 0.915 \end{tabular} 
& \begin{tabular}{@{}c@{}} 0.973 \\ 0.957 \end{tabular}\\ 

\hline
\midrule
Dev & \begin{tabular}{@{}c@{}} BN \\ AdaBN \end{tabular} 
& \begin{tabular}{@{}c@{}} 0.979 \\ 0.927 \end{tabular} 
& \begin{tabular}{@{}c@{}} 0.819 \\ 0.8 \end{tabular} 
& \begin{tabular}{@{}c@{}} 0.994 \\ 0.963 \end{tabular} 
& \begin{tabular}{@{}c@{}} 0.81 \\ 0.873 \end{tabular} 
& \begin{tabular}{@{}c@{}} 0.385 \\ 0.538 \end{tabular} 
& \begin{tabular}{@{}c@{}} 0.0 \\ 0.333 \end{tabular} 
& \begin{tabular}{@{}c@{}} 0.686 \\ 0.766 \end{tabular} 
& \begin{tabular}{@{}c@{}} 0.931 \\ 0.904 \end{tabular} \\ 

\hline
\midrule
Test & \begin{tabular}{@{}c@{}} BN \\ AdaBN \end{tabular} 
& \begin{tabular}{@{}c@{}} 0.984 \\  0.953 \end{tabular} 
& \begin{tabular}{@{}c@{}} 0.871 \\  0.847 \end{tabular} 
& \begin{tabular}{@{}c@{}} 0.964 \\  0.918 \end{tabular} 
& \begin{tabular}{@{}c@{}} 0.892 \\ 0.903 \end{tabular} 
& \begin{tabular}{@{}c@{}} 0.176 \\ 0.353 \end{tabular} 
& \begin{tabular}{@{}c@{}} 0.0 \\  0.0 \end{tabular} 
& \begin{tabular}{@{}c@{}} 0.683 \\  0.715 \end{tabular} 
& \begin{tabular}{@{}c@{}} 0.940 \\  0.914 \end{tabular}\\ 

\bottomrule
\end{tabular}
}
\end{center}
\vspace{-0.1cm}
\end{table*}

\subsection{Implementation details}

The style classification model is adapted from \cite{Emotionmultimodal} and is shown in Figure \ref{fig:Styleframework}. Specifically, we set the batch normalization layer with 0.9 momentum to help cross-domain adaptation. To compensate for the imbalance among style labels, we weighted the by-class loss function and the per-class accuracy with an inverse of style label prior and capping the neutral label prior to 0.25. Besides, AdaBN \cite{li2018adaptive} is implemented in this model to boost domain adaptation performance between the TTS and multi-style datasets. 

The multi-style TTS system is trained using the commissioned TTS dataset with style embedding features as conditional input features. The style embedding labels were generated by passing each utterance through the style classification model, as described in Section \ref{styleModel}. In the synthesis phase, the style embedding features could be automatically extracted from the input query or manually assigned as a combination of different styles.   



\subsection{Style classification}
\label{ssec:stylec}

In the style classification task, we first tested the style classifier model performance on the IEMOCAP train/test split. It achieves an overall accuracy of 72.7\%, which is similar to the reported state-of-the-art \cite{Emotionmultimodal}. 
To improve the embedding quality on the TTS dataset, the IEMOCAP dataset and the labeled subset of the TTS dataset were combined during training. The results show that the style classifier achieves 91.4\% overall accuracy and 71.5\% weighted accuracy on the TTS labeled dataset. 
With a lack of labeled data in anger and sadness in the TTS dataset, the prediction accuracy of these two classes is not high. The style classification accuracy decreased slightly on the IEMOCAP dataset after joint training, likely due to the mismatch between the TTS and IEMOCAP datasets. 

We performed normalization on the input features. The normalization is performed corpus-wise to compensate for the domain difference between our TTS dataset and the IEMOCAP dataset. Table \ref{tab:features} shows that normalizing both MFCC and prosody provides the best classification accuracy on the TTS dataset's validation set. So in the final model, we normalized both MFCC features and prosody features. The final classification accuracy for the TTS dataset is in Table \ref{tab:results}.

\begin{table*}[h]
\vspace{-0.2 cm}
\caption{TTS data F0 statistics: the Happy style has higher mean F0 than other styles. And the F0 standard deviations of Angry, Happy and Sad are larger than Neutral, Rushed and Soft styles.}
\vspace{-0.6 cm}
\label{tab:F0}

\begin{center}
\scalebox{1}{
\begin{tabular}{lcccccc}
\toprule
{Style}
 & {Angry} & {Happy} & {Sad} & {Neutral} & {Rushed} & {Soft} \\

\hline
\midrule
F0 & \small{195.5$\pm$30.8}
& 214.8$\pm$37.3 & 197.3$\pm$30.8 & 183.7$\pm$10.3 & 181.9$\pm$12.8 & 180.5$\pm$14.7 \\ 






\bottomrule
\end{tabular}
}
\end{center}
\vspace{-0.3cm}
\end{table*}

\begin{table}[t]
\vspace{-0.3 cm}
\caption{Feature selection: Normalizing MFCC and prosody vector can improve the performance of style classifier.}
\label{tab:features}
\vspace{-0.6cm}
\begin{center}

\scalebox{1}{
\begin{tabular}{lcc}
\toprule
\multirow{2}{*}{Features} & \multicolumn{2}{c}{{\bf Accuracy}} \\
 & {Weighted} & {Unweighted} \\

\hline
\midrule
Unnormalized & 0.726 
&  0.875 \\ 

\midrule
Normalized MFCC 
& 0.673
& 0.840\\ 

\midrule
Normalized prosody 
& 0.494 
& 0.62\\ 

\midrule
Normalized both 
& 0.766
& 0.904 \\ 

\bottomrule
\end{tabular}
}

\end{center}
\end{table}

\subsection{Multi-style TTS with conditional style embedding}
\label{ssec:subhead}

To evaluate our multi-style TTS's performance, we collected subjective evaluation responses from 22 listeners.
As reported in \cite{IE2018learning, banse1996acoustic, scherer1991vocal}, the human perception on the emotions of natural speech is only around 50\%, showing the ambiguity of emotion perception. Hence, instead of evaluating the subjective style accuracy on the multi-style synthesis results, we conducted the ABX test and preference test. Synthesis samples of our system are available at \cite{styleDemo}.

\subsubsection{ABX test}
The ABX test is designed to evaluate whether two styles generated with the same style embedding are perceived to be closer in speaking style when compared to a sample with a different style embedding. Since the style embedding can be used as a probability distribution over the 6 styles, to synthesize audio in a certain style, we construct the style embedding vector to have a value of 0.95 for the selected style and 0.01 for the other five styles. We designed the ABX test as follows. Given two different styles, we randomly choose an example in each style. We denote these two examples as $A$ and $B$. We then randomly choose a different sample $X$ from one of these two styles as reference. We then ask the listener to listen to samples $A$, $B$, and $X$, and then select which of $A$ or $B$ is perceived to be of the same style as the reference $X$. 

We created $15$ test sets in total, each of which corresponds to a pair of styles $A$ and $B$, and a reference $X$ of which the linguistic content is emotionally neutral. $22$ listeners participated in the test, which gives a total of $330$ ABX test comparison scores. We achieved an overall accuracy of $82.42\%$ (i.e. total number of matching pairs divided by the total number of ABX tests), indicating that the multi-style TTS is able to generate samples with perceivably distinguishable styles. 

\subsubsection{Preference test}

The preference test is designed to compare TTS responses generated by a default TTS without multi-style capability and the multi-style TTS when the style embedding is explicitly provided. Specifically, we ask the listeners to choose between TTS responses synthesized with the same text but different models: baseline TTS model (i.e., TTS without style embedding) or the multi-style TTS model. For the multi-style TTS responses, we provide the utterance with style of either the neutral style or, when appropriate, a hand-crafted style embedding (i.e., other style) based on the style of the text, assigned as a soft probability label whose style weights are determined based on the content of the utterance.

Results in Table \ref{tab:subjective} show that the multi-style TTS is preferred over the baseline TTS 72\% of the time, indicating strong user-preference when an appropriately styled TTS response is provided. It is interesting to note that the neutral style from the multi-style TTS is preferred by the listeners most of the time. This is largely due to the content of the test utterances, which is best spoken with a peaceful and relaxing neutral style. This result is consistent with the findings in \cite{IE2018learning}, which states that listeners prefer appropriate variation over random variation.


\begin{table}[!t]
\vspace{-0.4 cm}
\caption{Subjective Preferences: the proposed TTS model's results are preferred over the baseline TTS model's.}
\label{tab:subjective}
\vspace{-0.6cm}
\begin{center}
\scalebox{1}{

\resizebox{\columnwidth}{!}{
\begin{tabular}{lccc}
\toprule
& \multirow{2}{*}{{\bf Baseline TTS}} & \multicolumn{2}{c}{{\bf Multi-style TTS}} \\
 & & {Neutral Style} & {Other Styles} \\

\hline
\midrule
Preference (\%) & 28.0
& 54.2 
& 17.8\\

\bottomrule
\end{tabular}
}
}
\end{center}
\vspace{-0.6cm}
\end{table}

\subsubsection{Mimicking real life input query with styled TTS response}
\label{realexp}
We conducted experiments to evaluate the generalization capacity of the close-loop style extraction and multi-style TTS system. We recorded speech queries from multiple speakers who have never been seen in the training of our framework. These speakers read the queries freely in a quiet conference room. We then generated TTS responses for each query by conditioning on its style embedding. Our results show that over 40\% of test pairs are evaluated as good matches by listeners. We noticed that when the speaking style of the input query is strong, the TTS response can match the input style to a certain extent (samples are at \cite{styleDemo}). This can potentially be improved with more coherent style labels between the style extraction model training data and the TTS dataset.

\section{Discussions}
\label{sec:print}


In our proposed system, the soft (probability) style embedding is a weighted representation of different styles such that increasing the weight of a certain style emphasizes that style's effect on the synthesis outputs, shown in \cite{styleDemo}. This demonstrates the multi-style TTS's capability of synthesizing styled-speech with respect to the soft style embedding. 
We noticed utterance-mean F0 differs for different styles in the synthesis results, representing the style difference. For example, the inference result of the Happy style has a significantly higher F0 mean than the other styles. This is consistent with the statistics of F0 for different predicted classes in TTS training data, as shown in Table \ref{tab:F0}. 

We also noticed that the the Happy style of multi-style TTS has a significantly higher F0 mean than the other styles. This could be due to the reason that the model focused on the most distinguishable feature, such as F0 mean and failed to learn the nuances of the F0 contour. To mitigate this problem, the F0 mean and the F0 contour can be modeled separately. 
In addition, the sad and angry styled audio quality was comparatively worse than other styles, which could be due to the lack of anger and sadness samples in the TTS dataset. In the future, the performance of the multi-style TTS system can be further improved with a training dataset that contains more balanced style labels.


\section{Conclusions}
\label{sec:page}
In conclusion, we attempted to develop a style-embedded TTS that is more contextual and interactive. As shown in Section \ref{ssec:subhead}, with perfect style embedding, the system generated preferred TTS responses compared to a single style TTS. With automatically extracted style embeddings from real speech queries, the system demonstrated moderate capability in mimicking the speaking style of the input speech query. The overall quality can be improved with a more balanced multi-style TTS dataset and more coherent style labels between the style extraction model training data and the TTS dataset. 


\bibliographystyle{IEEEtran}

\bibliography{mybib}


\end{document}